\documentclass[aps,prl,twocolumn,showpacs,preprintnumbers,amsmath,amssymb]{revtex4-1}

\usepackage{graphicx}
\usepackage{bm}
\usepackage{upgreek}

\begin{document}

\title{{\em In-situ} transport characterization of magnetic states in Nb/Co Superconductor/Ferromagnet
heterostructures}

\author{Olena M. Kapran$^1$, Roman Morari$^{2,3}$, Taras Golod$^1$, Evgenii A. Borodianskyi$^1$, Vladimir Boian$^2$, Andrei Prepelita$^2$, Nikolay Klenov$^{4,5}$, Anatoli Sidorenko$^{2,6,7}$ and Vladimir M. Krasnov $^{1,3}$}
\email[E-mail: ]{Vladimir.Krasnov@fysik.su.se}

\affiliation{$^1$ Department of Physics, Stockholm University,
AlbaNova University Center, SE-10691 Stockholm, Sweden;}

\affiliation{$^2$ Institute of Electronic Engineering and
Nanotechnologies, MD2028 Chisinau, Moldova; }

\affiliation{$^3$ Moscow Institute of Physics and Technology,
State University, 141700 Dolgoprudny, Russia;}

\affiliation{$^4$ Lomonosov Moscow State University Skobeltsyn
Institute of Nuclear Physics, Moscow, 119991, Russia;}

\affiliation{$^5$ Moscow Technical University of Communication and
Informatics, 111024 Moscow, Russia;}

\affiliation{$^6$ Laboratory of Functional Nanostructures, Orel
State University named after I.S. Turgenev, 302026, Russia;}

\affiliation{$^7$ Technical University of Moldova, MD2004,
Chisinau, Moldova.}

\begin{abstract}
Employment of the non-trivial proximity effect in
Superconductor/Ferromagnet (S/F) heterostructures for creation of
novel superconducting devices requires an accurate control of
magnetic states in complex thin-film multilayers. In this work we
study experimentally in-plane transport properties of
micro-structured Nb/Co multilayers. We apply various transport
characterization techniques, including magnetoresistance, Hall
effect and the first-order-reversal-curves (FORC) analysis. We
demonstrate how FORC can be used for detailed {\em in-situ}
characterization of magnetic states. It reveals that upon
reduction of external field magnetization in ferromagnetic layers
first rotates in a coherent scissor-like manner, then switches
abruptly into the antiparallel state and after that splits into
the polydomain state, which gradually turns into the opposite
parallel state. The polydomain state is manifested by a profound
enhancement of resistance caused by flux-flow phenomenon,
triggered by domain stray fields. The scissor state represents the
noncollinear magnetic state in which the unconventional
odd-frequency spin-triplet order parameter should appear. The
non-hystertic nature of this state allows reversible tuning of the
magnetic orientation. Thus, we identify the range of parameters
and the procedure for {\em in-situ} control of devices based on
S/F heterostructures.

\end{abstract}



\maketitle

\section{I. Introduction}

Competition between spin-polarized ferromagnetism and spin-singlet
superconductivity leads to a variety of interesting phenomena
including possibile generation of the odd-frequency
spin-triplet order parameter \cite{Buzdin1999,Kadigrobov2001,Bergeret2001}.
In recent years this exotic state has been extensively studied
both theoretically
\cite{Buzdin2005,Efetov2005,Fominov,Blanter2004,Eschrig,Houzet_2007,Golubov,Trifunovic_2011,Melnikov_2012,Pugach_2012,Linder_2012,Richard_2015,Hikino_2015,Linder_2015,Ren_2016}
and experimentally
\cite{Bell_2004,Robinson_2010,Leksin_2011,Zdravkov_2013,Iovan_2014,Dresselhaus_2014,Ovsyannikov,Robinson_Sc2010,Khaire_2010,Banerjee_2014,Iovan_2014,Glick_2017,Aarts_2017,Martinez_2016,Strunk_2017,Sidorenko_2016,Lenk_2017,Kapran_2020}
in various Superconductor/Ferromagnet (S/F) heterostructures. It
is anticipated, that this phenomenon can be employed for creation
of novel superconducting devices, in which supercurrent is
determined and controlled by the magnetic state of the
heterostructure, i.e., by the relative orientation of
magnetizations in several F-layers
\cite{Bell_2004,Robinson_2010,Iovan_2014,Dresselhaus_2014,Robinson_Sc2010,Khaire_2010,Banerjee_2014,Iovan_2014,Glick_2017,Aarts_2017,Martinez_2016,Strunk_2017,Sidorenko_2016,Lenk_2017,Kapran_2020,Moodera_2019}.

However, practical realization of such devices is complicated
because neither ways of controlling many degrees of freedom in S/F
multilayers, nor methods for monitoring magnetic states in micro-
or nano-scale S/F devices are established. The situation is
complicated by a variety of coexisting phenomena: (i) Both singlet
and triplet currents with short and long-range components can flow
through S/F heterostructures \cite{Melnikov_2012}. Therefore, even
long-range supercurrent can not be automatically ascribed to the
triplet order. (ii) Supercurrent strongly depends on a usually
unknown domain structure in F
\cite{Iovan_2014,Aarts_2017,Weides_2008}, flux quantization in S
\cite{Golovchanskiy_2016,Iovan_2017}, both influenced by size and
geometry. (iii) The long-range spin-triplet supercurrent appears
only in the noncollinear magnetic state
\cite{Efetov2005,Houzet_2007,Golubov,Trifunovic_2011,Melnikov_2012}.
Therefore, utilization of this phenomenon for device applications
requires accurate determination and control of the micromagnetic
state of micro- or nano-scale devices. A similar control is needed
for operation of a large number of superconducting spintronics
devices, including memory elements and spin valves
\cite{Qader_2014,Bakurskiy_2016,Bakurskiy_2018,Shafraniuk_2019,Klenov_2019,Dresselhaus_2014,Kapran_2020,Glick_2017}.
The need for establishing experimental characterization techniques
for {\em in-situ} monitoring of magnetic states in S/F micro and
nano-devices is our main motivation.

Here we study experimentally in-plane transport properties of
micro-structured Nb/Co multilayers (ML's) with different number of
layers and layer thicknesses. Our goal is to demonstrate how
conventional experimental techniques can be used for {\em in-situ}
assessment of magnetic states of small S/F devices. The key
technique that we employ is the first-order-reversal-curves (FORC)
analysis. We demonstrate that in combination with
magnetoresistance (MR) and Hall effect measurements, it can
provide a detailed knowledge of the magnetic configuration in the
ML. In particular we identify the parallel (P), the antiparallel
(AP), the noncollinear monodomain scissor-state and polydomain
states. We observe that the domain state is manifested by a
profound enhancement of resistance. Analysis of the Hall effect
reveals that those maxima are associated with flux-flow
phenomenon, caused by motion of Abrikosov vortices induced by
domain stray fields. From device application perspective, the most
important is the noncollinear scissor state, in which the
unconventional odd-frequency spin-triplet order parameter should
appear. The non-hystertic nature of this state allows reversible
tuning of the magnetic configuration. Thus, we identify the range
of parameters and the procedure for controllable operation of
devices based on S/F heterostructures.

\begin{figure*}[t]
    \centering
    \includegraphics[width=0.95\textwidth]{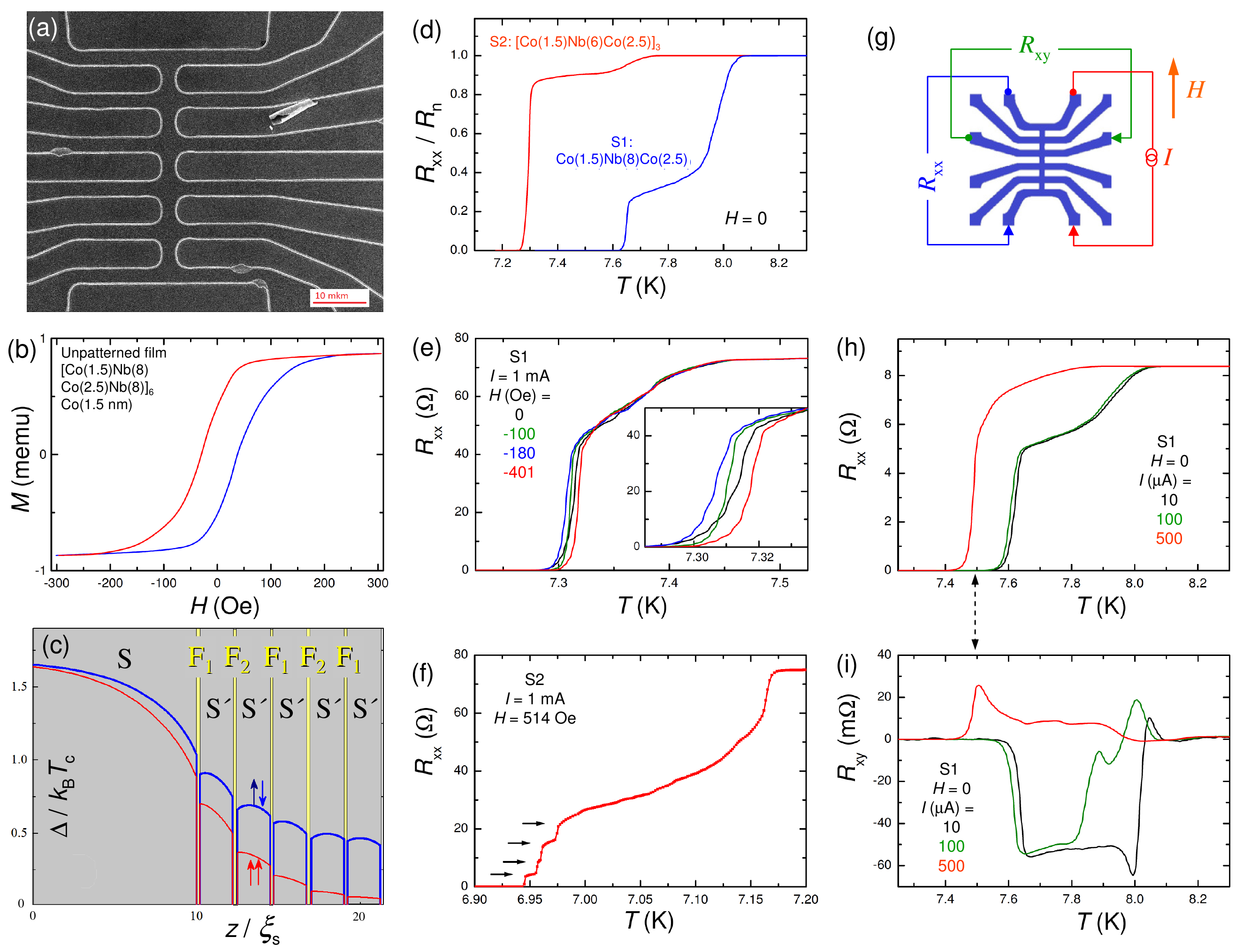}
    \caption{(color online).
    (a) SEM image of a micro-patterned Nb/Co multilayer. The sample contains 12 contacts, six horizontal and
    one vertical bridge with the widths of few microns.
    (b) Magnetization curve of an unpatterned Nb(25)[Co(1.5)Nb(8)Co(2.5)Nb(8)]$_6$Co(1.5 nm)Nb(25) ML. Blue/red curves represent up/down
    field sweeps (data from Ref.\cite{Klenov_2019}).
    (c) Simulated superconducting energy gap, $\Delta$, in the S2 ML for P (red)
    and AP (blue) states.
    (d) Temperature dependencies of longitudinal resistances, normalized by $R_n(T \gtrsim T_c)$, for
    both ML's. Measurements are taken with small currents at $H=0$. Double transitions are attributed to
    different critical temperatures of thick bottom S-layer and thin S'-spacers.
    (e) $R_{xx}(T)$ for a horizontal bridge (S1) at four consecutively increasing fields and $I_{ac}=1$ mA.
    Inset demonstrates a nonmonotonous dependence on magnetic field.
    (f) $R_{xx}(T)$ for a horizontal bridge (S2) at $H=514$ Oe and $I_{ac}=1$ mA.
    The resistive transition is significantly broadened, compared to
    the case in (d). Several small steps (marked by arrows) with similar resistance increments are seen.
    (g) Contact configuration for measurement of longitudinal and Hall resistances for (h) and (i) and the
    orientation of magnetic field in all experiments.
    (h) Longitudinal and (i) Hall resistances for a vertical bridge at S1 sample at different bias currents and $H=0$.
    A significant and sign-reversal Hall signal is observed within the transition
    region. This is a fingerprint of a flux-flow phenomenon,
    caused by motion of Abrikosov vortices in S-layers.
} \label{fig:fig1}
\end{figure*}


\section{II. Samples}

We study two types of Nb/Co ML's with different number of layers
and layer thicknesses. The simplest S1: Nb(50 nm)/Co(1.5 nm)/Nb(8
nm)/Co(2.5 nm)/Nb(8 nm)/Si ML (bottom -to -top), has just two
dissimilar Co layers composing a single pseudo spin valve. A more
complex S2: Nb(50 nm)/[Co(1.5 nm)/Nb(6 nm)/Co(2.5 nm)/Nb(6
mn)]$_3$Co(1.5 nm)/Nb(6 nm)/Si (the structure in square brackets
is repeated three times) has five Co layers. ML's are deposited by
magnetron sputtering in a single deposition cycle without breaking
vacuum. We use Nb target (99.95 \% purity) for deposition of
S-layers, Co (99.95 \% purity) for F-layers, and Si (99.999 \%)
for seeding bottom and protective top layers. ML's are grown on a
(1 1 1) Si wafers. Prior to deposition, targets were precleaned by
plasma-etching for 3 minutes and in addition for 1 minute upon
switching between targets. The deposition is performed at room
temperature with water cooled sample stage. Thicknesses are
defined using calibrated growth rates: 3.5 nm/s for Nb and 0.1
nm/s for Co. For every set of layers, three identical samples were
prepared simultaneously, of which some were used for calibration
of films etching rates. ML's are patterned into micron-scale
bridges with multiple contacts using photolithography and reactive
ion etching. Scanning electron microscope (SEM) image of one of
the studied samples is shown in Fig. \ref{fig:fig1} (a).

Control of the magnetic state implies a possibility of variation
of a relative magnetization orientation in neighbor F-layers,
which requires different coercive fields. To facilitate this we
use dissimilar Co layers with thicknesses 1.5 and 2.5 nm. Nb/Co
ML's with similar Co thicknesses have been studied earlier and
demonstrated good uniformity and perspectives for device
applications
\cite{Robinson_Sc2010,Khaire_2010,Martinez_2016,Sidorenko_2016,Lenk_2017,Klenov_2019,Robinson_2020}.
Fig. \ref{fig:fig1} (b) shows a magnetization curve for a similar
Nb(25 nm)/[Co(1.5 nm)/Nb(8 nm)/Co(2.5 nm)/Nb(8 nm)]$_6$Co(1.5
nm)/Nb(25 nm) unpatterned ML film, deposited using the same setup
(data from Ref. \cite{Klenov_2019}). $M(H)$ is obtained by SQUID
magnetometer in field parallel to the film in the normal state,
$T>T_c$. A significant hysteresis of $M(H)$ reveals the in-plane
anisotropy of Co films (albeit with a small coercive field,
$H_C\sim 30$ Oe), consistent with earlier studies
\cite{Co_Parkin_1990,Co_Mosca_1991,Co_Broeder_1991,Robinson_2020}.

Fig. \ref{fig:fig1} (c) shows numerical simulation of the
superconducting order parameter, $\Delta$, distribution in S/F ML
similar to S2 \cite{Note3}. It provides a qualitative
understanding of modulation of the proximity effect in the ML in P
(red) and AP (blue) states. A thick bottom S-layer Nb(50 nm) acts
as a Cooper pair reservoir and is only modestly affected by
F-layer orientation. However, thin S' spacers, Nb(6 and 8 nm),
with the thickness comparable to the superconducting coherence
length, $\xi_S\sim 10$ nm, are strongly affected.
Superconductivity in S' layers is stronger in the AP-state and is
almost quenched in the outmost S' layer in the P-state. This
demonstrates the tunability of superconductivity in such S/F ML's
by changing the magnetic state. Because of the bottom S-layer,
there is a gradient of $\Delta$ in S' layers, which implies that
S' layers have dissimilar superconducting properties.

Measurements are performed in a closed-cycle $^3$He cryostat with
a superconducting magnet. More details about fabrication,
characterization and experimental setup can be found elsewhere
\cite{Klenov_2019,Iovan_2014,Kapran_2020}. Resistances are
measured by the lock-in technique with different current
amplitudes $I_{ac}$. In all cases magnetic field is applied
parallel to the film plane in the orientation, sketched in Fig.
\ref{fig:fig1} (g). Multi-terminal geometry of samples allows
simultaneous four-probe measurements of different segments of the
sample in both longitudinal, $R_{xx}$, and Hall, $R_{xy}$,
directions. When current is sent through the central vertical
bridge, as sketched in Fig. \ref{fig:fig1} (g), measurements
correspond to the easy-axis magnetization orientation (field along
the long side of the vertical line). Alternatively we can send
current through horizontal bridges, which corresponds to the
hard-axis magnetization orientation (field perpendicular to the
long side of the bridge).

\begin{figure*}[t]
    \centering
    \includegraphics[width=0.99\textwidth]{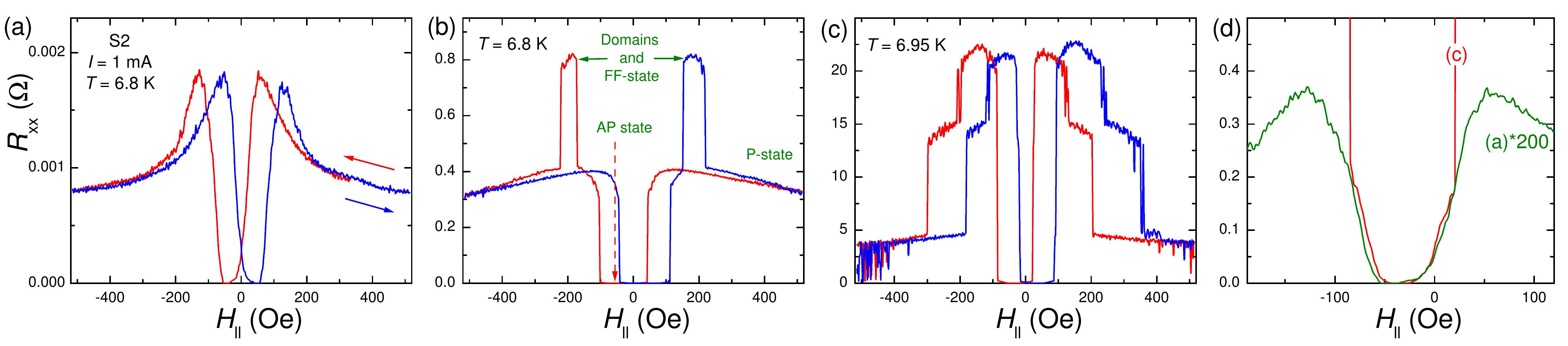}
    \caption{(color online).
    Longitudinal magnetoresistances for a horizontal bridge at S2, measured at different
    $T$.
    (a) At the lowest $T\simeq 6.8$ K, corresponding to the
    very onset of resistivity.
    (b) At marginally higher $T$. Here an additional maximum
    appears, which is attributed to triggering of the flux-flow
    phenomenon by domain.
    (c) At $T\simeq 6.95$ K, corresponding to the middle of the resistive transition, shown in Fig.
    \ref{fig:fig1} (f). Steps are attributed to transitions of individual spacer
    layers.
    (d) Low-resistance parts of the curves from (a) and (c), which demonstrate a similar behavior in the AP-state.
} \label{fig:fig2}
\end{figure*}

\section{III. Results}

Fig. \ref{fig:fig1} (d) shows $R_{xx}(T)$ dependencies, normalized
by the normal state resistance $R_n(T>T_c)$, for micro-bridges at
S1 (blue) and S2 (red) ML's at $H=0$. Resistances are measured
with $I_{ac}=10~\mu$A for S1 and $20~\mu$A for S2, which
correspond to approximately equal small current densities in both
ML's. Both ML's show a double transition, which could be
attributed to different critical temperatures in S and S' layers,
$T_c'(S')<T_c(S)$. Consistent with this assumption, $T_c'$ of the
S2 ML with thinner S'(6 nm) is smaller than for S1 ML with S'(8
nm).

Fig. \ref{fig:fig1} (e) shows $R_{xx}(T)$ curves for a horizontal
bridge at the S1 sample at four sequentially increasing magnetic
fields (hard-axis orientation) and $I_{ac}=1$ mA. It is seen that
the onset of resistivity at $T\sim 7.3$ K is affected by the
field. However, this effect is nonmonotonous with field, as can be
seen from the zoomed-in view in the inset. The rest of transition
is little affected in this field range $\lesssim 500$ Oe. This is
caused by the large value of the upper critical field for thin Nb
films in parallel field \cite{Zeinali_2016}. Therefore, the
observed nonmonotonous field dependence is not directly induced by
the applied field but reflects the remagnetization process of
F-layers in the ML.

Fig. \ref{fig:fig1} (f) shows the $R_{xx}(T)$ curve for a
horizontal bridge at the S2 sample (hard axis orientation),
measured at $H=514$ Oe with $I_{ac}=1$ mA. Here several steps
(marked by arrows) with similar resistance increments $\sim
5~\Omega$ can be distinguished close to the onset of $R_{xx}(T)$.
They are probably due to individual transitions of five S' layers
in this ML, which have a gradient of the order parameter, as seen
from Fig. \ref{fig:fig1} (c).

\subsection{III A. Non-linear flux-flow Hall effect}

Resistivity in type-II superconductors with sizes significantly
larger than the London penetration depth, $\lambda$, is caused by
motion of Abrikosov vortices, i.e., has a flux-flow (FF) nature
\cite{Samoilov_1995,Krasnov_1997,Sonin_1997,Zeldov_2017}. Since
our micron-size bridges are significantly larger than $\lambda\sim
100$ nm of Nb, the FF phenomenon is anticipated. FF depends on
magnetic field, transport current and vortex pinning, which
depends on temperature. For a single S-layer at $H=0$ vortices may
only be induced by the self-field of transport current. However,
for S/F heterostructures they may also be induced by stray fields
from F-layers, especially in the presence of domain walls
\cite{Sonin_2003,Vlasko_2008,Aladushkin_2009,Cucolo_2012}. Domain
walls also affect the FF phenomenon because they create a pinning
landscape for vortices: vortices are pinned to domains and can not
move across them, but can freely move along domain walls
\cite{Vlasko_2008}. Therefore, we expect that the magnetic state
of a ML may influence FF resistance.

Fig. \ref{fig:fig1} (h) shows $R_{xx}(T)$ for a vertical bridge at
S1 sample with contact configuration depicted in Fig.
\ref{fig:fig1} (g). All the curves are obtained at $H=0$, without
remagnetizing the ML. However, measurements are made with
different current amplitudes $I_{ac}$: $10~\mu$A (black),
$100~\mu$A (olive) and $500~\mu$A (red). At low and intermediate
bias currents, 10 and $100~\mu$A, a two-step $R_{xx}(T)$
transition occurs with $T_c(S)\simeq 8$ K and $T_c'(S')\simeq 7.8$
K and the shape of $R_{xx}(T)$ is remaining the same. This
indicates that the current-voltage characteristics at low bias is
almost linear and resistance is almost bias independent. However,
at high bias, $I_{ac}=500~\mu$A, the $R_{xx}(T)$ is significantly
smeared out and the onset shifts to significantly lower $T$. This
manifests entrance into the non-linear regime.

Fig. \ref{fig:fig1} (i) shows corresponding Hall resistances
measured as shown in Fig. \ref{fig:fig1} (g). It is seen that a
significant Hall signal appears only within the resistive
transition region \cite{Note1} and is non-linear (bias-dependent)
even at low bias. Furthermore, at low and intermediate bias
$R_{xy}$ changes sign. Such a behavior is typical for the FF Hall
effect in superconductors
\cite{Samoilov_1995,Krasnov_1997,Sonin_1997}. This provides clear
evidence for FF phenomenon in our samples. From comparison of
Figs. \ref{fig:fig1} (h) and (i) it is clear that FF takes place
in the whole resistive transition region.

\subsection{III B. Flux-flow magnetoresistance, triggered by domains}

Figure \ref{fig:fig2} shows longitudinal magnetoresistances at
different temperatures. Measurements are done on a horizontal
bridge (hard axis orientation) at the S2 sample, the same as in
Fig. \ref{fig:fig1} (f), and at the same current $I_{ac}=1$ mA.
Blue/red lines represent up/down field sweeps. The curves in Fig.
\ref{fig:fig2} (a) are measured at the very onset of the resistive
transition, as seen from small (m$\Omega$) resistance values. Here
we observe the simplest MR curves, most consistent with the
two-state theoretical prediction based on mono-domain simulations,
as in Fig. \ref{fig:fig1} (c). Namely, P state at high field with
large $R_{xx}$ (suppressed superconductivity) and AP state at low
fields with small $R_{xx}$ (enhanced superconductivity)
\cite{Buzdin2005,Efetov2005,Fominov,Eschrig,Houzet_2007,Golubov,Moodera_2019,Leksin_2011,Sidorenko_2016,Kapran_2020}.
However, this simple scenario does not explain appearance of
additional maxima between P and AP sates. The discrepancy becomes
more pronounced at higher $T$, i.e., higher up at the $R_{xx}(T)$
transition.

Fig. \ref{fig:fig2} (c) shows MR at almost the same $T$ as in (a)
with a higher, but still quite small $R_{xx}<1~\Omega$. Fig.
\ref{fig:fig2} (c) represents MR at higher $T$, corresponding to
the middle of the first transition in $R_{xx}(T)$ from Fig.
\ref{fig:fig1} (f). Here the maxima between P-state (high fields)
and AP state ($R_{xx}\simeq 0$ at low fields) become profound.
Appearance of such a maximum also follows from the non-monotonous
field variation of $R_{xx}(T)$ curves, as can be seen from the
inset in Fig. \ref{fig:fig1} (e). In Fig. \ref{fig:fig2} (c) it is
seen that the resistance switches stepwise between certain values
$\sim$ 0, 5, 10 (missing in (c), but accessible at slightly
different $T$), 15 and 20 $\Omega$. They correspond to steps in
$R_{xx}(T)$ indicated in Fig. \ref{fig:fig1} (f), which we
attributed to transition of individual S' layers. From Figs.
\ref{fig:fig1} (b) and (c) it can be seen that the fields at which
resistance drops slightly depends on $T$. This does not allow a
straightforward association of such the drop with transition into
the AP state, which should be $T$-independent in this $T$-range.

In Fig. \ref{fig:fig1} (d) we replot low-field parts of the
downward curves from (a) and (c). It is seen that they coincide
after proper scaling, i.e. they represent $T$-independent part of
MR, which could be associated with the magnetic state. In this
case the minima with $R_{xx}\simeq 0$ from -20 to -50 Oe should
represent the range of existence of the AP state.
Observation of this scaling points out that the underlying
magnetic state is independent of $T$ and the difference in shapes
of MR is caused by something else. The origin of the observed
unusual three-state MR with and additional profound maximum
between AP and P states is not difficult to deduce.
The third magnetic state is very well known - the polydomain
state. It for sure appears in micron-scale F-films. Since the
resistivity of thin S-films has FF origin, the observed unusual
maximum in MR is likely caused by triggering of FF phenomenon by
domain stray fields, which change upon remagnetization of
F-layers. The remarkable temperature variation of MR in Figs.
\ref{fig:fig2} (a-c) is then primarily caused by temperature
variation of vortex mobility.

\begin{figure}[t]
    \centering
    \includegraphics[width=0.49\textwidth]{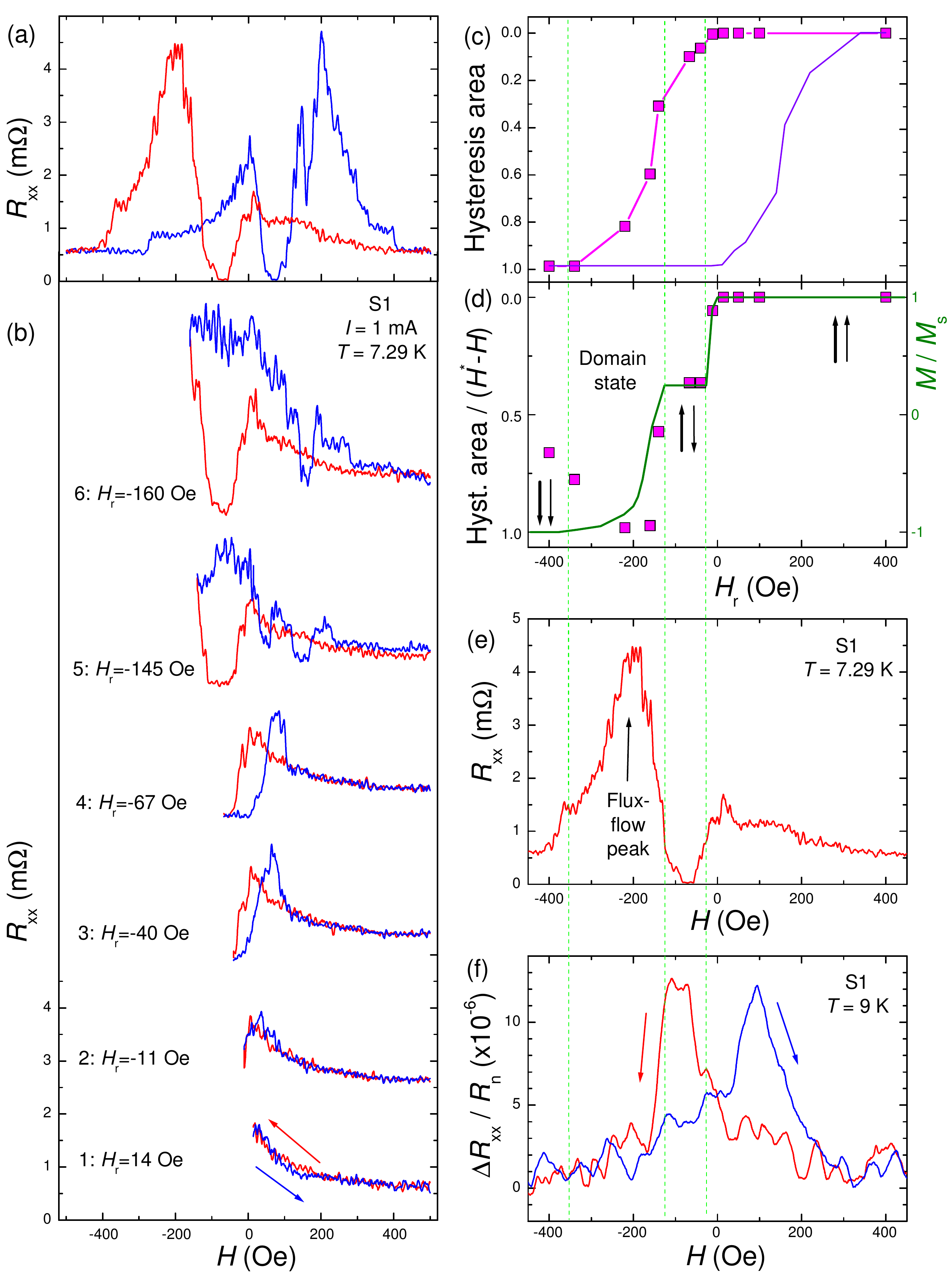}
    \caption{(color online). (a) Magnetoresistance of a horizontal bridge on the S1 sample at $T=7.29$ K.
    (b) FORC analysis of MR for different reversal fields $H_r$ (curves are shifted vertically for
    clarity).
    (c) Hysteresis area between forward (red) and reversal (blue)
    curves in (b). Violet curve shows a central-symmetric
    reflection of the same data.
    (d) Hysteresis area normalized by the integration field range $H^*-H_r$, where $H^*$ is the field for onset of the hysteresis.
    Olive line (right axis) represents a sketch of the expected magnetization curve.
    (e) Full-scale $R_{xx}(H)$ curve at $T=7.29$ K for a downward field sweep.
    (f) Normal state magnetoresistance $(R_{xx}(H)-R_n)/R_n$ for the same bridge at $T=9$
    K. Green vertical lines emphasize correlations between
    features in (c-f).
}
    \label{fig:fig3}
\end{figure}

\subsection{III C. {\em In-situ} characterization of magnetic states by the first-order-reversal-curves analysis}

The aim of this work is to establish techniques for {\em in-situ}
transport characterization of magnetic states in micro-patterned
S/F ML's. The key technique that we employ for this purpose is
FORC - a powerful tool for characterization of magnetic states in
complex ferromagnetic structures
\cite{FORC1_2008,FORC2_2013,Akerman_2014}. FORC analysis starts at
the same saturated state. Then field is swept to a reversal field
$H_{r}$ and measurements are carried out on the way back to the
saturated state. The experiment is repeated with gradually varying
$H_{r}$. Recently it has been shown that FORC can be used for the
analysis of magnetic states in S/F ML's \cite{Kapran_2020}.
Essentially we have to search for appearance of different types of
hysteresis in the FORC response, which indicate switching into
some specific metastable magnetic states. There are two mechanisms
for appearance of hysteresis in the ML's \cite{Iovan_2017,
Kapran_2020}. The major hysteresis is associated with switching in
and out from the magnetostatically stable AP state. Multiple
smaller ones are associated with switching between different
domain states, which are also metastable.

Fig. \ref{fig:fig3} (a) shows longitudinal MR for the same
horizontal bridge (hard axis orientation) at S1 sample as in Fig.
\ref{fig:fig1} (e). The overall shape of $R_{xx}(H)$ with a
minimum at low fields and an additional FF maximum nearby is
similar to that for S2 in Fig. \ref{fig:fig2} (b). Fig.
\ref{fig:fig3} (b) represents FORC analysis for this bridge.
Magnetic field is swept from above the saturation field $H \simeq
+500$ Oe down to the reversal field $H_r$ and back to the
saturation field. Red lines represent forward curves and blue -
the FORC's. For $H_r\simeq +14$ Oe, curve-1 in Fig. \ref{fig:fig3}
(b), the FORC is reversible. The non-hysteretic behavior
corresponds to coherent monodomain rotation of magnetization in a
scissor-like manner \cite{Iovan_2017, Kapran_2020}. At $H_r = -11$
Oe, curve-2, the FORC starts to show a tiny signature of
hysteresis, which disappears at $H>50$ Oe. For $H_r\simeq -40$ Oe,
curve-3, the forward (red) curve reached the minimum $R_{xx}\simeq
0$, characteristic for the AP state, and the reversal curve (blue)
start to exhibit a clear hysteresis. With further increase of
$H_r$ within the minimum the reversal curve is practically
unchanged, as seen from the curve-4. This indicates that the state
of the ML remains the same. As reported in Ref. \cite{Kapran_2020}
appearance of the initial hysteresis at small fields is associated
with switching to the AP state. This is fully consistent with our
observation that the initial hysteresis corresponds to the minimum
of $R_{xx}$.

With further increase of $H_r$, beyond the AP minimum, the
reversal curve clearly changes, see curve-5, indicating switching
into a different magnetic state. Furthermore the range of
hysteresis expands to $H>200$ Oe and several additional small
switches occur within this range. The curve-6 in Fig.
\ref{fig:fig3} (b) is obtained for $H_r=-160$ Oe in the middle of
the $R_{xx}$ maximum. The reversal curve is clearly different from
curve-5 revealing yet another initial state. With further increase
of $H_r$ towards the negative saturation field, the hysteresis
changes gradually until reaching the saturated state, shown in
Fig. \ref{fig:fig3} (a). Such a gradual transformation indicates
that the ML is in a polydomain state with many close metastable
states.

For a more quantitative analysis of FORC data, following Ref.
\cite{Kapran_2020}, in Fig. \ref{fig:fig3} (c) we plot
$H_r$-dependence of normalized hysteresis area between forward and
reverse curves, i.e. the integral of the absolute value of the
difference between red and blue curves in Fig. \ref{fig:fig3} (b).
Magenta symbols represent FORC data.
They are plotted in a reverse scale along the vertical axis to
resemble magnetization curves. From this plot it is clearly seen
how the hysteresis start to develop at $H_r\lesssim -10$ Oe and
saturates at $H_r\lesssim -350$ Oe. However, this curve does not
represent the magnetization curve. In particular, at the AP
minimum in the range -120 Oe$\lesssim H_r\lesssim$-20 Oe the state
of the ML remains the same, as follows from the similarity of FORC
curves 3 and 4 in Fig. \ref{fig:fig3} (b). In this range the
hysteresis area is growing linearly with $H_r$ simply because the
integration range is increasing as $H^*-H_r$, where $H^*$ is the
field for onset of hysteresis. In Fig. \ref{fig:fig3} (d) we plot
hysteresis area divided by $H^*-H_r$. With such a normalization
the AP state is properly described by a plateau. For comparison we
also show a sketch of the expected magnetization curve (olive
line, right axis), with the AP plateau at 0.25=(2.5nm
-1.5nm)/(2.5nm+1.5nm) of the saturation magnetization $M_s$. It is
seen that such representation provides a remarkably close
description of $M(H)$ curves at intermediate fields. At higher
fields $H<-220$ Oe the points start to fall down, which indicates
approaching to the saturation state at which the hysteresis area
is no longer depending on the integration range $H^*-H_r$, as seen
from Fig. \ref{fig:fig3} (c).

For completeness of the analysis, in Figs. \ref{fig:fig3} (e) we
replot the full MR curve and in Fig.\ref{fig:fig3} (f) show the
normal-state MR for the same bridge at $T=9$ K. The two rightmost
vertical lines in (c-f) marks the onset and the end of the AP
state. The leftmost vertical line marks the onset of the
saturation. The two rightmost lines emphasize a clear correlation
between the onset of hysteresis (c,d), the minimum of resistance
in the superconducting state (e) and the maximum of resistance in
the normal state (f), which all are signatures of the AP state.
The two leftmost vertical lines indicate correlations between the
FF maximum in $R_{xx}$ (e) and a gradual transition state between
the AP state and the negative P-state (d,f).

Thus, from FORC analysis we conclude that upon remagnetization of
the ML from the P-state, it first enters into a coherently
rotating, nonhysteretic, scissor state, after which it abruptly
switches into the AP state, stays in it for a while and then
breaks into a polydomain state, which triggers the flux-flow
phenomenon. With further increase of field the polydomain state
gradually turns into the opposite P-state. This picture is
consistent with the assessment based on Hall effect, Fig.
\ref{fig:fig1} (i), and MR analysis, Fig. \ref{fig:fig2}, and with
earlier FORC analysis of Ni-based SF$_1$NF$_2$S spin valves
\cite{Kapran_2020}.

\section{IV. Discussion}

As we've seen, information about evolution of magnetic states in
S/F ML's is encoded in the shapes of MR curves. The most
remarkable feature of those is the three-state remagnetization
with an unusual maximum in addition to well understood P and AP
states. As we already mentioned, the third state has to be the
well known polydomain state. Below we substantiate this statement.

First, we note that resistance in our films has the flux-flow
origin, as unambiguously shown by Hall effect measurements, Fig.
\ref{fig:fig1} (i). Therefore, MR is due to modulation of FF. The
latter depends on the vortex density, pinning, the superconducting
order parameter and the driving current. Next, we can do the
following exclusion:

(i) Vortices in thin films have a pancake structure (Pearl
vortices) with field perpendicular to the film. Therefore, vortex
density depends on magnetic induction $B_z$ perpendicular to the
film. Since the applied field in our experiment is parallel to the
film, it does not contribute to $B_z$.

(ii) Parallel magnetic field does suppress superconductivity.
However, it should cause a monotonous (parabolic) increase of
$R_{xx}$ with increasing field, which is not the case. Thus, field
variation as such does not explain the observed nonmonotonous MR.

(iii) Since all measurements in Fig. \ref{fig:fig2} are done with
the same $I_{ac}=1$ mA, the current self-field in not changing
and, therefore, can not cause modulation of MR.

The above exclusion leaves one possible source of MR specific for
S/F ML's: variation of magnetic state in F-layers. However,
remagnetization of F-layers can contribute to flux-flow phenomenon
only if it generates the perpendicular field component. Since our
Co layers have the in-plane anisotropy, this is only possible via
stray fields from domain walls
\cite{Sonin_2003,Vlasko_2008,Aladushkin_2009,Cucolo_2012}. This
brings us to the following interpretation of the observed unusual
MR, as indicated in Fig. \ref{fig:fig2} (b): MR modulation has two
contributions: First is the conventional suppression/enhancement
of superconductivity (order parameter modulation) in the P/AP
states at high/small fields
\cite{Buzdin2005,Efetov2005,Fominov,Eschrig,Houzet_2007,Golubov,Moodera_2019,Leksin_2011,Sidorenko_2016,Kapran_2020}.
The second contribution is caused by the flux-flow phenomena
triggered by domains. Domains both create vortices and generate a
pinning landscape that determines the direction of vortex motion
\cite{Vlasko_2008}. Domains could explain the observed dramatic
variation of MR shapes with $T$: at low $T$, Fig. \ref{fig:fig2}
(a), FF is negligible (vortices are almost immobile) and the MR is
dominated by order parameter modulation, leading to appearance of
a minimum at low fields. With increasing $T$ vortices get depinned
and FF modulated by domain texture becomes a dominant factor,
causing appearance of additional maxima in-between P and AP states
as in Figs. \ref{fig:fig2} (b,c).

Generally, the polydomain state is unwanted in most devices based
on S/F heterostructures because it is hard to control. Domains
cause an irreversible behavior of the heterostructure, associated
with both the magnetic hysteresis and with generation of Abrikosov
vortices, which are pinned at film defects. As shown previously
\cite{Iovan_2014}, those two factors may dramatically distort
characteristics of S/F devices.

FORC is the key technique, that we are advertising for {\em
in-situ} transport characterization of magnetic states. As shown
in sec. III C and in Ref. \cite{Kapran_2020}, FORC may provide a
detailed information about variation of magnetic states in micro
and nano-scale S/F devices. In Ref. \cite{Kapran_2020} this was
demonstrated using out-of-plane measurements. Here we show that a
similar information can be obtained using in-plain transport
measurements. From FORC analysis in Figs. \ref{fig:fig3} we can
characterize evolution of micromagnetic states in our structures
upon remagnetization. We observe that the initial stage of
remagnetization from P to AP state is fully reversible, see
curve-1 in Figs. \ref{fig:fig3} (b). Micromagnetic simulations
show that such stage corresponds to monodomain coherent rotation
of magnetizations in neighbor F-layers in opposite directions in a
scissor-like manner \cite{Iovan_2017,Kapran_2020}. As seen from
Fig. \ref{fig:fig3} (d), the hysteresis appears abruptly upon
switching into the AP-state. The AP-state is magnetostatically
stable and, therefore persists in a certain field range. However,
switching from AP to the opposite P-state occurs gradually, which
according to micromagnetic simulations indicate splitting into
polydomain state \cite{Iovan_2017,Kapran_2020}. The more domains -
the more gradual is the transition to the saturated state.
Appearance of domains triggers the flux-flow state by introducing
Abrikosov vortices, as reported before
\cite{Vlasko_2008,Aladushkin_2009,Cucolo_2012} and leads to
appearance of the additional FF-maximum in MR.

\section{Conclusions}

To conclude, we have studied in-plane transport properties of
micro-structured Nb/Co multilayers. We demonstrated how
conventional transport techniques can be used for assessment of
magnetic states of small S/F heterostructures and devices. For
this we apply various experimental techniques, including
magnetoresistance, Hall effect and first-order-reversal-curves
analysis. We have shown that a combination of those techniques,
performed simultaneously, can provide a detailed knowledge about
evolution of micro-magnetic states. FORC is the key technique that
we advertise for such {\em in-situ} characterization. Using it we
identified the parallel, the antiparallel, the mono-domain
scissor-state and polydomain states. Polydomain states are
manifested by a profound enhancement of resistance cause by
flux-flow phenomenon, triggered by domain stray fields.

Importantly, the scissor state corresponds to the noncollinear
magnetic state of the multilayer in which the unconventional
odd-frequency spin-triplet order parameter should appear in the
heterostructure. The non-hystertic nature of this state allows
controllable tuning of magnetic orientation. Thus, we identify the
range of parameters and the procedure for controllable operation
of superconducting spintronic devices based on S/F
heterostructures. Essentially we conclude that for moderately
small (micrometer-scale) devices controllable and highly
reversible operation can be achieved at fields between one of the
P-states down to the AP state without entering into the realm of
the opposite P state.

\subsection{Acknowledgments}
We are grateful to Sergey Bakurskiy, Igor Soloviev, Andrey
Schegolev, Yury Khaydukov, Mikhail Kupriyanov and Alexander
Golubov for stimulating discussions. The work was supported by the
European Union H2020-WIDESPREAD-05-2017-Twinning project
``SPINTECH" under grant agreement Nr. 810144 (sample preparation
and low temperature measurements), the Russian Science Foundation
grant No. 19-19-00594 (V.M.K.: data analysis and manuscript
preparation), and partially by the project STCU {\#6329} "Full
switching memory element for spintronics on the base of
superconducting spin-valve effect" (A.S.). The manuscript was
written during a sabbatical semester of V.M.K. at MIPT, supported
by the Faculty of Natural Sciences at SU and the Russian Ministry
of Education and Science within the program "5top100".



\end{document}